# A stochastic model for tumor heterogeneity

Giuseppina Simone

*University of Naples Federico II, Piazzale Tecchio 80, 80125 Napoli, Italy*

*Email: giuseppina.simone@unina.it*

**Abstract.**

Phenotype variations define heterogeneity of biological and molecular systems, which play a crucial role in several mechanisms. *Heterogeneity* has been demonstrated in tumor cells.

Here, samples from blood of patients affected from colon tumor were analyzed and fished with a microfluidic assay based on galactose active moieties, and incubated – for culturing – in SCID mice. Following the experimental investigation, a model based on Markov theory was implemented and discussed to explain the equilibrium existing between phenotypes of subpopulations of cells sorted using the microfluidic assay.

The model in combination with the experimental results had many implications for tumor heterogeneity. It displayed interconversion of phenotypes, as observed after experiments.

The interconversion generates of metastatic cells and implies that targeting the CTCs will be not an efficient method to prevent tumor recurrence. Most importantly, understanding the transitions between cell phenotypes in cell population can boost the understanding of tumor generation and growth.

Biological systems, including in the mammalian cells and bacteria, during their life can be subject to variation of the phenotype *(Aird, 2007; Meacham et al., 2013)*. In the last two decades, the researchers explained this phenomenon making fundamental assumptions of biological molecule fluctuations (Zhou et al., 2005). Fluctuations result from stochastic noise in biochemical reactions, or can be manifested as transitions between different biological phenotypes (McDonnell et al., 2009). In general, biochemical reactions, which describe the molecular recognition between the cells and binding molecules, arise stochasticity from intermolecular collisions.

Phenotype variations define heterogeneity of biological and molecular systems, which play a crucial role in several mechanisms (Marusyk et al. 2010). Importantly, *heterogeneity* has been demonstrated in tumor (Heppner, 1984). Stochastic fluctuations are typical in 'small' numbers (Gillespie, 1977); however, it has been shown that dramatic molecular fluctuations can occur even when the copy numbers of some biomolecular components are large.

Here, in order to understand heterogeneity in tumor cells, a microfluidic assay was used 'to read cell phenotype'. The results were analyzed using a stochastic approach.

Previously, it was observed that the epithelial circulating cells included several subpopulations. Focusing the attention on the role that the molecular association between the galectin, coating the membrane of tumor cells, and the $\beta$-galactose immobilized on the bottom layer of a microfluidic assay, it was observed heterogeneity of epithelial cells circulating in human blood (Simone, 2013). The protocols were shown and discussed elsewhere (Simone, 2013; Simone, 2012). Summarizing, blood samples from patients affected from colon tumor were analyzed. In particular, at the beginning, according to the heterogeneous size and relative cell density of cancer cells, a density-ring between 1080−1090 g mL$^{-1}$ was collected. The cells, in the range of densities collected, were, prevalently, epithelial. Indeed, the first analysis was carried out on CD45 positive/negative cell filtration, by selecting for the scope of the investigation CD45$^+$ cell subpopulation (epithelial cells).

The collected cells were analyzed at the cytofluorimeter and it was observed that the cell membrane was coated by galectin, a protein that forms specific molecular association with the $\beta$-isomer of the galactose.

The cells in analysis were, finally, analyzed with the microfluidic assay, consisting in galactose (to bind the galectin) active moieties able to record different cell fingerprints.

The impact of the research carried out with the microfluidics and the glycans included that the galectin fished positive/negative subpopulations and that the populations of Circulating Cells – identified in patients affected by tumor – included subpopulations displaying different phenotypes. The results here discussed refer to investigations involving colon tumor – affected patients; however, heterogeneity was observed in breast cancer cells, as well (Simone, 2014).

The plot in *Figure 1a* displays the composition of the 'native' and 'unbound' cell population. Indeed, the cells were labeled as 'bound', such as the cells that adhered to the galactose microfluidic assay, 'unbound', such as the cells that were recovered during the step of washing of the assay and 'native', such as the cells after the initial purification.

The plot in Figure 1a, displaying the results of a cytofluorimetric investigation on the cell receptors, refers to galectin positive/negative and cytokeratin-20 positive/negative (widely expressed in colon cancer cells). During the experimental investigation, it was observed that two subpopulations, with galectin positive phenotype, were recognized, one $CK20^+$ and one $CK20^-$.

Analyzing the results, it was observed that $gal^+/CK^+$ experienced a significant reduction at the exit of the assay, many cells with this phenotype were captured by the galactose moieties; $gal^-/CK^-$ passed through the assay, but they were collected at the exit of the assay as an intact subpopulation; $gal^-/CK^+$ cells, which expressed cytokeratin positive phenotype but have galectin negative phenotype, passed through the assay without variation. It worth to mention here that the subpopulation displaying $gal^-/CK^+$ phenotype was a tiny portion of the whole population, however, the most important results was that $gal^+/CK^+$ cells represented the metastatic portion of the circulating cells (Simone, 2013). To date, the attention was focused on them. Finally, it worth to mention that the cells selected for the investigation did not display a subpopulation with phenotype $gal^+/CK^-$.

The experiments showed that the $gal^+/CK^+$ cells were identified as Circulating Tumor Cells (CTCs) and the rest of the subpopulations, which was recovered at the exit of the microfluidic assay, were identified as Residual Epithelial Cells (RHCs).

Before introducing the samples inside the assay, the native samples from human blood were filtered, in order to isolate epithelial cells (Malara, 2014) for a subpopulation representing the middle 15% of the native distribution. The population initially purified was incubated inside the microfluidic assay and sorted, galectin +/- was the cut-off. The sorting using a filter galectin positive a narrow band of the cell population was isolated. RHC and CTC cells were sorted into their different phenotypes to >70 % purity, according with early and advanced colon tumor (**Figure 1b**). Here, the first addressed question was whether there was inter-conversion between the cell phenotypes, as it was expected looking at stem tumor cells.

In order to address the question, the culture of the cells was taken in consideration. Indeed, the cells, galectin positive (CTCs) and negative (RHCs), after microfluidic investigation, were inoculated in SCID mice and observed after 70-80 days. Furthermore, the experimental model indicated that both *bound* and *unbound* cells (else CTCs and RHCs) could rise tumor cells, in order to reestablish the initial phenotype equilibrium. To test this hypothesis, the sorted cell subpopulations were injected into SCID. The first experiment displayed that only the bound cells fraction produced tumors and tumor cells circulating in mouse blood. However, a hypothesis was that the predominately unbound subpopulation did not survive long enough to re-establish population equilibrium. To date, the cells were injected in Matrigel suspension to enhance cell viability. Finally, the new experiment displayed that all three subpopulations generate tumors.

Cytofluorimetric analysis of human CK20 (a cytokeratin specific for labeling tumor colon cells) on cell suspension from brain and spleen of mice displayed that during the time of culturing, the sorted cell subpopulations rapidly converged towards the phenotype equilibriums that were seen in the native cells (**Figure 1c and 1d**). Indeed, the bound cells, which before the culture in mice displayed a cytofluorimeter readout (Figure 1c) after the culture time generated in mice cells with the expression of galectin and cytokeratin (Figure 1d). However, unbound cells, which were considered galectin-free and potentially non-tumor, surviving in mice, were able to generate the metastatic phenotype.

Speculation about the results display that, after the incubation, the proliferation rates between each cell phenotype did not markedly differ, and, the cells that were in a minority would need to divide rapidly to re-establish equilibrium galectin-cytokeratin receptors. To date, it can be made hypothesis that interconversion between cell phenotypes is probable.

In order to address the first question, a mathematical model was implemented.

The model, based on the studies of the possible cell transitions, assumed that the conversion probabilities of the cells depended on its current state and not its prior states.

Stochastic approaches can explain the existence of different phenotypes within the cells, which is defined as 'dynamic heterogeneity' or interconversion between phenotypes.

Dynamic heterogeneity is described by Markov theory (Gupta, 2011). Markov theory and model accounted for transition processes between states which are memory-free. Such 'states' can refer to populations of predators and prey or cell phenotype or more in general as ecological habitats.

**Figure 2a** displays the Markov model schematization and a possible cell interconversion and the different states. The cell might reach the state of replication, mortality, phenotype conversion, or maintenance.

Here, a Markov model has been used to describe the generation of heterogeneity. This kind of model is traditionally presented using the set of differential equations in (1) and (2).

$$\frac{dC1}{dt} = c_{C_1}C1 + (r_{C_2} - m_{C_2} - c_{C_2})C2 \qquad (1)$$

$$\frac{dC2}{dt} = c_{C_2}C2 + (r_{C_1} - m_{C_1} - c_{C_1})C1 \qquad (2)$$

A cell from heterogeneous tumor population, C1, replicates with rate coefficient r, dies (mortality) with rate coefficient m, and converts to the C2 state with rate coefficient c. A cell from the C2 population has analogous rate coefficients r, m, and c, leading to the dynamics for C2 and C1.

The model has been used to critically discuss the experimental findings above displayed, where C1 and C2 can be considered the unbound or RHC cells and the bound or CTC cells. According to this fundamental memory-free hypothesis of Markov model and the Equation (1) and (2), at time t, the cell might eventually fall into slots as displayed in Figure 2a. The mathematical equations have been solved according to methods displayed in (Wilkinson, 2009).

Stochasticity is synonymous of randomness; indeed, any event is a collection of them that can be also not distributed around a single most probable value. Intrinsic stochasticity is generated by the dynamics of the system from the random timing of individual reactions. Low number of molecules for which are more significant the changes leads intrinsic stochasticity. According to this, running the model a number of times, different trajectories were collected for each simulation (**Figure 2b**). Comparing the results, RHC cells display a wider spread compared with the CTCs, evidencing that the *good cells, the RHCs,* fit better with the stochastic behavior and they have deeper propensity to experience a *change of status*. The ensemble statistic is displayed in panels **2c, d** of **Figure 2**, the mean and the standard deviation for the species, respectively. However, from these plots it appears that RHC and CTC cells do not have exactly identical standard deviations (Figure 2d), though the variation of their means is different (Figure 2c).

The stochastic simulation algorithm displayed that the reaction systems is not conservative (results of the simulation are not shown), that explain why the standard deviations are different. In order to summarize the results, the variation of the mean and standard deviation for the CTCs were plotted (**Figure 2e**). From the plot, it can be observed that CTCs started out with a zero value at the beginning, but both its mean and variance approached non-zero very sharply and stayed there.

The model in combination with the experimental results had many implications for tumor heterogeneity. It displayed interconversion of phenotypes, as observed after experiments.

The interconversion generates of metastatic cells and implies that targeting the CTCs will be not an efficient method to prevent tumor recurrence. Most importantly, understanding the transitions between cell phenotypes in cell population can boost the understanding of tumor generation and growth.

*The author declares to do have no competing interests.*

**Acknowledgements.**
The author wants to thank Dr. N. Malara for interesting discussions on cell behavior, Circulating Tumor Cells, and cytofluorimetric technique.

**References.**

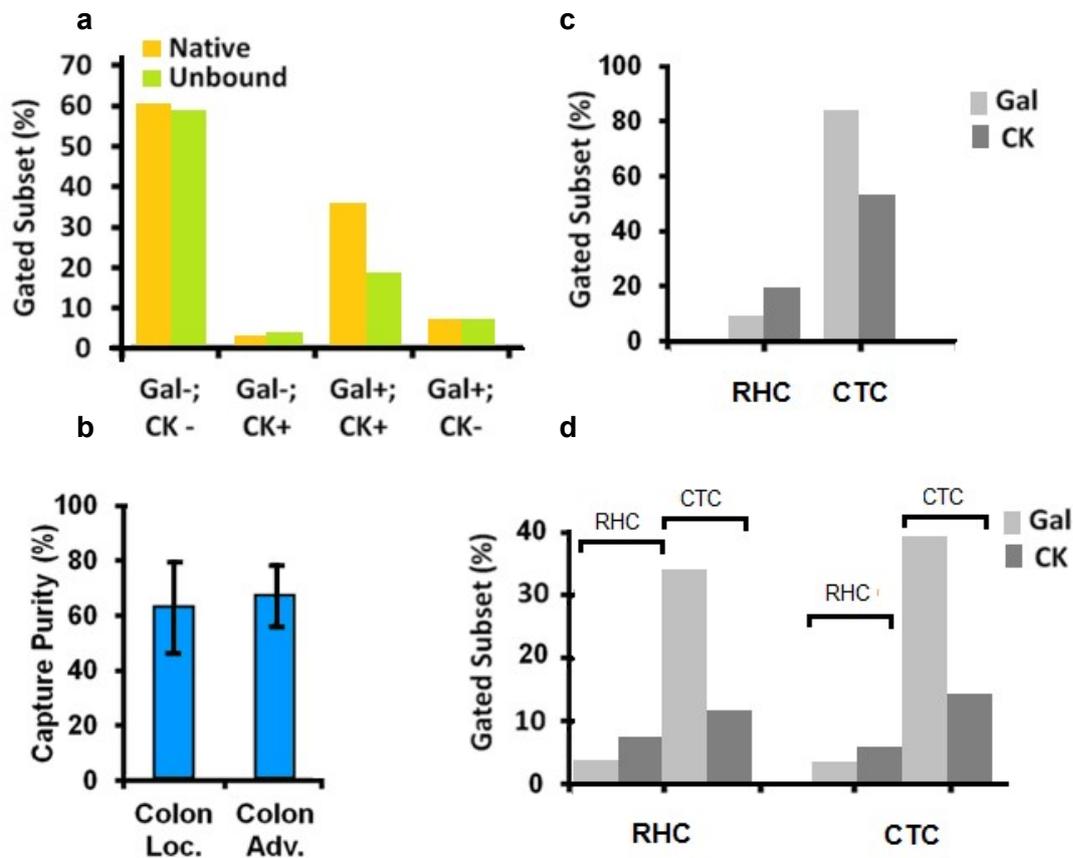

Figure 1. a). Cytofluorimetric analysis of expression of galectin-3 and cytokeratin in colon cancer samples from blood of patients. The native population and the unbound cells are heterogeneously composed by galectin + and galectin- and by cytokeratin + and cytokeratin-. The fraction of gal+ CK+ was identified as CTCs (epithelial cells) the fraction of gal+ CK- was identified as RHCs. The incubation inside the assay reduces the fraction of CTCs at the outlet. The increase at the outlet of gal+CK- includes the presence of beads and debris of cells. b) Purity of the assay purity. c) Proportions of cells in native colon cells from tumor-affected patients. d) Cellular subpopulations in RHCs and CTCs were isolated by FACS with antibodies directed against the CK20 and galectin cell surface antigens. Bar chart show the proportion of cells in each cell-differentiation state as assessed by FACS after incubation in SCID mice for 70-80 days.

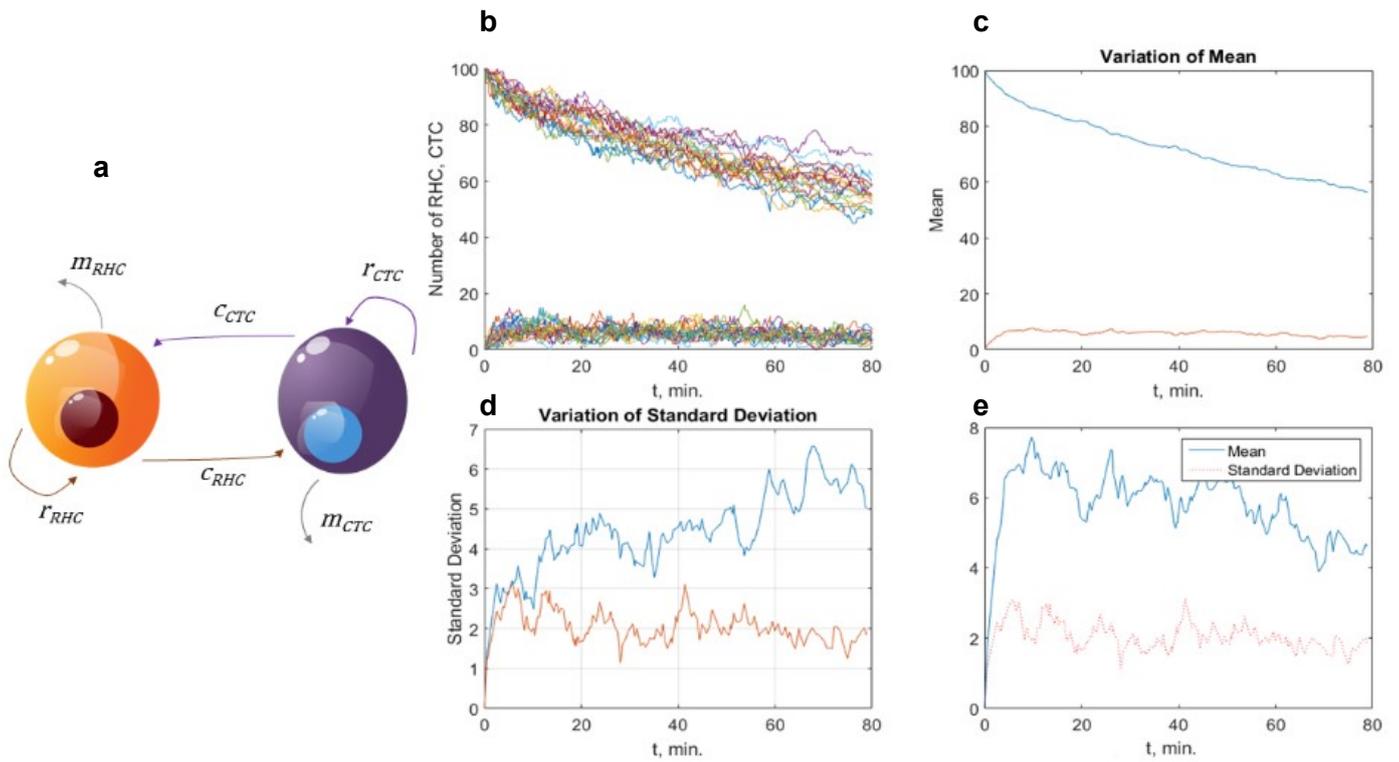

Figure 2. a). Cartoon to schematize Markov model. b) N-loop of the stochastic model. Profile of number of RHCs and CTCs. c) Variation of Mean based on curves of b). d) Variation of Standard Deviation based on curves b). e) Variation of the mean and standard deviation for the CTCs .